\documentclass{ijmpi_authors}

\usepackage{lipsum}

\begin{document}

\twocolumn[
\title{Schemes of Propagation Models and Source Estimators for Rumor Source Detection in Online Social Networks: A Short Survey of a Decade of Research}

\author{Jin}{Rong}{a,\ast}
%\author{Jin}{Rong}{}
\author{Wu}{Weili}{a}

\affiliation{}{Department of Computer Science, The Univeristy of Texas at Dallas, Richardson, USA}
%\affiliation{b}{Example Labs}
\affiliation{}{\{rong.jin, weiliwu\}@utdallas.edu}
\affiliation{\ast}{Corresponding author}

\maketitle

\begin{abstract}
Recent years have seen various rumor diffusion models being assumed in detection of rumor source research of the online social network. Diffusion model is arguably considered as a very important and challengeable factor for source detection in networks but it is less studied. This paper provides an overview of three representative schemes of Independent Cascade-based, Epidemic-based, and Learning-based to model the patterns of rumor propagation as well as three major schemes of estimators for rumor sources since its inception a decade ago. 
\end{abstract}
]
% =======================================================================

\section{Introduction}
\begin{figure}[!htb]
	\centering
	\includegraphics[scale=0.22]{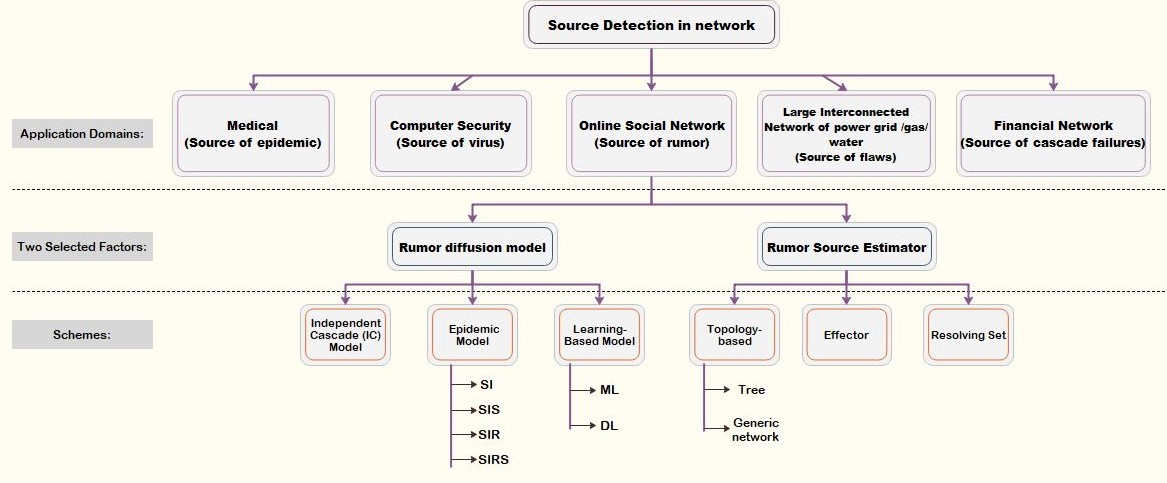}
	\caption{Roadmap of this short survey.}
	\label{fig:test}
\end{figure}

Source detection plays a vital role in any network like the Network of power grid, Network of peoples, Online Social Networks, etc. where the source identification has been performed to find the origin or source, such as, detecting the source of epidemics to control infection spreading \cite{PhysRevLett.114.248701}, finding the source of a computer virus in a network \cite{Shah2010}, locating gas leakage source in wireless sensor network \cite{Gasleakage}, identifying propagation sources in complex networks \cite{PhysRevE, Jiang2017}, investigating the sources of misinformation in online social networks \cite{DTNguyen2012}, and rumor source detection \cite{Shah2011RumorsIA, Shah2012FindingRS} in online social media network. From the existing survey work, such as \cite{Jiang2017} reviews source identification methods in accordance with three categories of network observation which is one of the major premises, \cite{MeiLi} surveys various information diffusion models based on two categories, and then a comprehensive study about factors to be considered for source detection of rumor in social network by \cite{SHELKE2019}. We observe that very little review has been done, from the perspectives of diffusion patterns and various estimator for sources, on rumor source detection problem in the online social networks. Different models refer to different application domains in seeking propagation origins. Therefore, the objective of this short survey paper is to summarize and discuss existing approaches to rumor source detection in views of these two aspects. The roadmap of this survey is depicted in Figure \ref{fig:test}. To the best of our knowledge, this is the first survey that focuses on the schemes of modeling rumor propagation and source estimation of seeking origins of rumor in online social networks.   

This short survey is structured as below. In Section II, we summarize existing three approaches to model rumor propagation and introduce basic mechanisms behind them. Section III shows existing three main schemes to estimate rumor sources. The followed evaluation metrics and datasets for rumor sources detection problem are briefly presented in Section IV. Research open issues and challenges in source detection of rumor are stated as a part of conclusion in Section V.

% (see~Figure~\ref{fig:test}).

% \begin{equation}
% 	\frac{d}{dx}\left( \int_{0}^{x} f(u)\,du\right)=f(x)\,.
% \end{equation}

% \begin{table}
% 	\centering
% 	\caption{This is another example.}
% 	\label{tab:test}	
% 	\begin{tabular}{c c c c c}
% 			 			&  A &  B &  C  &  D	\\
% 		 Test 1	& 0.00  & 1.35  &	0.27 & 0.44	\\
% 	     Test 2	& 0.00  & 1.35  &	0.27 & 0.44
% 	\end{tabular}
% \end{table}

\section{Rumor Propagation Model}
Modeling how information spreads is utilized to analyze how misinformation spread as well as to stop the spread of the misinformation such as rumors \cite{Guille2013}. In this section, we present basic knowledge of major rumor spreading models and summarize how these models have been employed by researchers to their rumor source(s) detection problems. We broadly classify models of rumor propagation into three categories: the Independent Cascade-based, epidemic-based and the state-of-the-art learning-based.
\subsection{Independent Cascade(IC) model}
\textit{Independent Cascade (IC) Model} is one of commonly used diffusion models \cite{KKT-ICmodel}. In IC model, the information passes in the network through cascades. It describes the process of information diffusion,  from a set of initially activated nodes named sources, proceeding on a directed graph where each node can be activated or not with a monotonicity assumption, i.e. activated nodes are the one who believed the rumor and become infected, and activated nodes cannot deactivate \cite{D'Angelo2016}. For IC model, when a node $v$ becomes active, it has a single chance of activating one of its currently inactive neighboring nodes, say a node $w$, with a numerical diffusion probability $p_{vw}$ associated to the edge between nodes $v$ and $w$. Researchers have assumed their diffusion process based on IC model to seek rumor source(s).

Xu \textit{et al}. \cite{Xuwen2015} firstly studied the problem of identifying single rumor source detection with respect to Independent Cascade (IC) model in online social networks when investigators have no relevant textual or content information. As cascades in IC model are necessarily trees, the influence structure of a cascade is given by a directed tree $T$, which is contained in the directed graph $G$. And, each cascade in IC model is associated with the propagation probability between any two neighboring nodes. Based on these two facts, Xu \textit{et al}. defined Maximum Influence Path and Maximum Propagation Tree as well as proposed a polynomial time rumor source detection algorithm under partial observations.     

Lim \textit{et al}. \cite{Lim2018} formulated and investigated a $k$-minimum distance rumor source detection problem, specially to find a small set of rumor candidates which can be used as initial seeds for further iterative query or investigation, by using an Independent Cascade (IC) model to analyze the likelihood of rumor sources. Since the propagation of rumor from one node to another usually incurs a certain amount of time delay, Lim \textit{et al}. modified the classic IC mdoel which has extra property that each edge $(u,v) \in E$ has associated a time-varying probability $p(u, v, t)$, characterizing that how much $u$ can influence $v$ at time $t$. 

\subsection{Epidemic based Model}
\textit{Epidemic Models} are the ways that describes the spreading, infection and recovery processes of diseases among the population. The basic epidemic models are used for finding the origin of viral disease \cite{MeiLi}. Similarly, they are employed by researchers for finding the sources of rumors in the online social networks according to different scenarios. So far, four classical types of epidemic models are mainly characterized in rumor diffusion as below. 

\begin{enumerate}
\item \textit{Susceptible-Infected (SI) Model}: In the SI model \cite{Pastor-Satorras}, there are two probable states of a node: susceptible (S) and infected (I). Nodes are initially susceptible and can be infected with spreading rumors. Susceptible nodes are uninfected nodes but having infected neighboring nodes. Once a susceptible node becomes infected due to contagion from its neighborhood, it remains infected forever. SI model only describes the infection process of $S$ $\rightarrow$ $I$ without taking into account that infected nodes can be recovered after having been infected. In terms of rumor diffusion in online social networks, infected nodes are those who have received or believed any rumor from their infected neighbors.

\item \textit{Susceptible-Infected-Susceptible (SIS) Model}: in the SIS model, the two states of susceptible (S) and infected (I) are the same as that in the SI model. But in this model, when susceptible nodes become infected, after some time they can yet again be susceptible after being cured. This infection and recovery process of $S$ $\rightarrow$ $I$ $\rightarrow$ $S$ addresses impracticality in SI model.   
    
\item \textit{Susceptible-Infected-Recovered (SIR) Model}: The SIR model divides the total population into three categories: S, I, and R (recovered), where S and I represent the susceptible and infected nodes as described in the previous two models. Different from SI or SIS model, an infected node in SIR model may be recovered, it will not spread the information and remains in the recovered state further \cite{Da2014}. The diffusion process can be characterized as $S$ $\rightarrow$ $I$ $\rightarrow$ $R$. When in a online social network, the recovered node is the one who recognizes of a rumor, therefore it will either delete/ignore the rumor message or not passing that rumor to neighbors.          

\item \textit{Susceptible-Infected-Recovered-Susceptible (SIRS) Model}: In SIRS model, it believes that a recovered node can become a susceptible node with probability $\alpha$ \cite{JIN2007}. Thus, the diffusion process in this model is 
$S \rightarrow I \rightarrow R \rightarrow S$.  

\end{enumerate}

Shah and Zaman \cite{Shah2011RumorsIA, Shah2016FRS} initially assume that the rumor propagation follows SI model originating from a single source. 
Dong \textit{et al}. \cite{Dong2013} extended study based on the setting of various suspect sources in the SI model. 
Luo \textit{et al}. \cite{Luo2013} later adopted the SI diffusion model for estimating multiple infection sources and their infection regions. 
Wang \textit{et al}. \cite{Wang2015} considered the value of diversity from multiple observations for single rumor source detection in SI spreading.
Choi \textit{et al}. \cite{Choi2017} studied the impact of querying in a highly
generalized setup for a rumor source detection under consideration of a SI rumor spreading model.  
Beside that, SIS model has been considered on inferring a single rumor source detection problem in Luo \textit{et al}. \cite{LuoT2013} and Wang \textit{et al}. \cite{Tan2015}. Luo \textit{et al}. derived an estimator based on estimating the most likely rumor source associated with the most likely infection path. Wang \textit{et al}. proposed a rumor centrality based algorithm that leverages multiple observations to first construct a diffusion tree graph, and then use their union rumor centrality to find the rumor source. SIR model was firstly integrated in detecting sources of computer viruses in networks by Shah and Zaman \cite{Shah2010}. After that, the spread of information in Zhu and Ying \cite{Zhu2016} also follows the SIR model for studying single information source detection problem in a network. Locating multiple sources in a networks under the SIR model has also been studied in Luo \textit{et al}. \cite{LuoT2012}, Zang \textit{et al}. \cite{ZANG2015} and Jiang \textit{et al}. \cite{Jiang2018}. So far, to best of my knowledge, SIRS model has not yet been applied in rumor source identification methods. Future work may take it into consideration.      

In information diffusion, there are also new models developed based on classical models such as SEIR (Suceptible Exposed Infected Recovered) model \cite{SEIR}, SEIRS (Suceptible Exposed Infected Recovered Susceptible) model \cite{SEIRS}, MSEIR (Passive-immunity Suceptible Exposed Infected Recovered) \cite{MSEIR}, etc. As far as we know, these models have not been applied in rumor source(s) detection problems by researchers.

\subsection{Learning based Model}
Since it is usually difficult to acquire the actual underlying propagation model in practice, many studies are proposed on an assumption that the underlying propagation model is known in advance and given as input. For instance, Independent Cascade(IC) model or one kind of Epidemic based models has been widely utilized in rumor source detection. However, we know that this assumption may lead to impracticability on real data and may limit their application range. Therefore, different types of machine learning methods \cite{MLonG} and deep learning methods \cite{DLonG} to graphs data have been studied and reviewed. In this short survey, we are only presenting learning based models in rumor source(s) detection problem in online social networks. 

Wang \textit{et al}. \cite{LPSI} firstly studies on multiple sources detection without knowing the underlying propagation model by proposing a Label Propagation based Source Identification (LPSI) method, which is based on the idea of source prominence as well as inspired by a label propagation based semi-supervised learning method \cite{Zhou03}. By setting an original label to the infection status of each node, positive labels (+1) to infected nodes and negative labels (-1) to uninfected nodes, LPSI lets these labels iteratively propagate in a snapshot of the network, and finally predicts local peaks of the convergent node labels as source nodes. Nonetheless, they still suffer from the shortcoming that the node label is simply an integer which may restrict the prediction precision. To improve prediction precision on the same problem, 
Dong \textit{et al}. \cite{Dong} then firstly attempts to apply the Graph Convolutional Networks (GCN) technique on multiple rumor source detection and then proposed a supervised learning based model named Graph Convolutional Networks based Source Identification (GCNSI). Basically, in GCNSI, Dong \textit{et al}. proposed an input generation algorithm to extend LPSI integer label into a multi-dimentional vector for each node as training data, and applied GCN in capturing different features of a node based on two assumptions, one is source prominence from LPSI, and the other is rumor centrality. Due to node representation utilizing its multi-order neighbors information by adopting spectral domain convolution so that their prediction precision on the sources is improved. Although these works are for multiple sources detection, they undoubtedly also fit to single source detection for sure.
More recently, a work \cite{sharma2020graph} also explored GCN for detecting possible suspicious users who are often involved in spreading the rumors on online social media. Additionally, another very recent work \cite{shah2020finding} revisited problem about locating the source of an epidemic, namely finding patient zero (P0), using Graph Neural Networks (GNNs) to learn P0. Similarly, we believe that it can be applied in rumor source(s) detection problem in online social networks.

To best of our knowledge, aforementioned works in this section are all existing studies of learning-based models regarding rumor source(s) detection problem in online social networks.

\subsection{Discussion}
% Create a table here to categorize the papers that develop rumor spreading models into these three categories.
Over the past decade of research on the methods of rumor source(s) detection, information propagation models and the network structures are taken into consideration. Works can be classified into two types: (1) infection status based
analysis \cite{Shah2011RumorsIA, ZANG2015}; (2) partial observations based analysis \cite{LuoTay, zejnilovic, SRS, Zhu2017CatchEmAL}. 
In most of the methods on the former type, it is a given input that the underlying propagation model is known in advance. Obviously, it is difficult to obtain the model information in practice. However, infection status based analysis has a broader application prospect since we cannot deploy observing nodes or acquire related information in some real-world situations. Those studies on learning based models have been carried out for source identification without the requirement
of knowing the underlying rumor propagation model.

Most of machine learning tasks on graphs - node classification, link prediction, learning over the whole graph, and community detection are very different from normal supervised/unsupervised learning. This is because graphs are interconnected with each other and data independence assumption fails. Researchers refer to it as semi-supervised learning.

Deep Learning techniques on graphs is the neural networks on graph data representation learning, which iteratively updates node representations by exchanging information between the neighbor nodes via relation paths and repeat the iteration until convergence and lets convolution layer capturing different features of a node. In the most cases, deep learning based models outperform among all baselines, but it spends time to train and tune parameters. Therefore, deep learning on graphs is non-trivial since several challenges exist in practical graphs such as irregular structures of graphs, diversity of graphs or large scale graphs and so on \cite{DLonG}.

\section{Rumor Source Estimator}
Rumor source estimator is a main challenge in rumor source detection problem. How to construct the rumor source estimator? In this section, we give an overview of source estimator that have been used for existing rumor source detection in online social networks. We classify schemes of estimators into three categories as shown in Figure \ref{fig:test}.

\subsection{Topology-based Estimator}
In this subsection, we summarize the rumor source estimator developed by the center of certain type of graph. There are two main types of graphs: tree-like networks and generic networks.

\textit{1) Regular Tree} Shah and Zaman \cite{Shah2010, Shah2011RumorsIA} firstly proposed rumor centrality as an estimator for a single source. They showed that the node with maximum rumor centrality (called rumor center) is the Maximum Likelihood Estimator of the rumor source if the underlying graph is a regular tree in social networks. 
After this publication, many other variants of this problem have been studied. 
Dong \textit{et al.} \cite{Dong2013} utilized the same definition of rumor center and proposed a notion of local rumor center as the node with the highest rumor centrality in the priori set of suspects to identify a single source for regular trees. 
Luo \textit{et al.} \cite{LuoT2013} proposed a multi-rumor-center method to identify multiple rumor sources in tree-structured networks. This method is too computationally expensive to be applied in large-scale networks as computational complexity of this method is $O(n^k)$, where $n$ is the number of infected nodes and $k$ is the number of sources. Zhu and Ying \cite{Zhu2016} proposed a Jordan center method, which utilizes a sample path based approach, to detect diffusion sources in tree networks with snapshot observations. Luo \textit{et al.} \cite{Luo2013} derived the Jordan center method using a different approach. Chen \textit{et al.} \cite{Chen2016} extended the Jordan center method from single source detection to the identification of multiple sources in tree networks. In addition, Shah and Zaman \cite{Shah2011RumorsIA} proved that, even in tree networks, the rumor source identification problem is a $\#P$-\textit{complete problem}, which is at least as hard as the corresponding NP problem.

\textit{2) Random Tree} Shah and Zaman \cite{Shah2016FRS} established the universality of rumor centrality for source detection for generic random trees without two limited settings of regular trees and the exponential spreading time set in \cite{Shah2010, Shah2011RumorsIA}, which inspired Fuchs and Yu \cite{Fuchs-Yu2015} extended work following the definition of the rumor center for grown simple families of random trees, which contain binary search trees, recursive trees and plane-oriented recursive trees.  

\textit{3) Generic Network} Shah and Zaman \cite{Shah2012FindingRS} extended their study in tree-like network to random graphs. Pinto \textit{et al.} \cite{Pinto2012} employ BFS technique to reconstruct generic networks into trees, and then the origin is sought in the BFS trees. 

\subsection{Effector Detection}
An activation state in a social network with a certain influence diffusion shows the users who have been influenced. The effectors are nodes that can best explain the observed activation state, which indicates that identification of effectors is important in understanding the dynamics of influence diffusion. Specially, we can interpret effectors as the critical nodes that trigger or stop the spread of information, or we may consider effectors as the source of the influence diffusion. Thus, effector detection problem \cite{Lappas2010} helps in identifying the source of rumors or infections.

Tong \textit{et al.} \cite{Tong} tackled the effector detection problem from a novel perspectives for social networks. Their approach is based on the influence distance that measures the chance that an active user can activate its neighbors. That is, for a certain pair of users, the shorter the influence distance, the higher
probability that one can activate the other. When we are given an activation
state, the effectors are expected to have short influence distance to active users while long to inactive users. By this idea, Tong \textit{et al.} proposed the influence-distance-based effector detection problem for IC model that was firstly studied in Lappas \textit{et al.} \cite{Lappas2010} and Tong \textit{et al.} provided a 3-approximation. In the problem, effectors can be played as rumor source estimators. However, For some activation states, it is challengeable to find best effectors from active nodes. The criteria for selecting effectors depend not only on the activation state but also on the diffusion model. Thus, it is interesting to investigate whether a meaningful effector exists for a given activation state.
 
\subsection{Resolving Set}         
In this section, we would like to present two novel rumor source(s) estimation approaches by Chen \textit{et al.} \cite{ChenWang} and Zhang \textit{et al.} \cite{SRS} in which they both study source detection problem from a deterministic point of view.

Chen \textit{et al.} firstly discovered that the concept of doubly resolving set (DRS) can be employed to study the single source detection problem and presented an $O(\ln n)$-approximation algorithm for the minimum weight DRS problem. 

After that, Zhang \textit{et al.} found that the other concept of set resolving set (SRS) can be applied for estimating multiple rumor sources independent of diffusion models in networks with partial observations based analysis. They let $G$ be a network on n nodes, a node subset $K$ is an SRS of $G$ if all detectable node sets are distinguishable by $K$. Then, the problem of multiple rumor sources detection in the network can be modeled as finding an SRS $K$ with the smallest cardinality. Zhang \textit{et al.} also gave a polynomial-time greedy algorithm for finding a minimum SRS in a general network with a performance ratio $O(\ln n)$. To best of our knowledge, there is currently no any other continued and expanded work to study these approaches on different diffusion models or different applications.  

\section{Evaluation}

\subsection{Evaluation Metrics}
An effective algorithm of finding a source of rumor spreading is a basic component of a rumor source detection system. In order to measure the performance of the algorithm, the choice of evaluation is usually measured by three different quality of localization metrics: the accuracy, the rank and the distance error. The accuracy is the empirical probability that a source found by the algorithm is the true source \cite{Luo2013}. The rank is the true source position on the nodes list, which is sorted in descending order by likelihood of being the source \cite{Xuwen2015}. The distance error is the shortest path distance between the real source and the source found by the algorithm \cite{Lim2018}.

\subsection{Datasets}
Datasets used for rumor source detection in social networks have been widely classified
into synthetic datasets and real-world datasets.

\textit{Synthetic datasets} are mainly structured in terms of tree
and graph. The Tree networks are represented by Random dregular
tree. Small-world (SW) networks and scale free networks
are basically used for graph network.

\textit{Real datasets} like Facebook, Twitter, Wiki-vote are
freely accessible on Stanford Large Network Dataset Collection. Similar to
Twitter, there is a popular Chinese micro blogging network, i.e. Sina
Weibo. The study of datasets used for rumor source detection problem in the social networks has been widely investigated in \cite{SHELKE2019}.

% \subsection{Comparative Study}
% In this section, we provide an overview of the performance of state-of-the-art systems on benchmark data sets.\\
% Create a table here and explain briefly.

\section{Conclusion}
The proliferation of data generated by a social network generates
a number of real-world problems to be solved, and rumor
source detection problem is one of them. This short survey aims to summarize and analyse existing schemes for modeling rumor diffusion process as well as estimating rumor source(s) in online social networks. In this paper, except for two oftenly reviewed influence diffusion model (i.e. IC model) and epidemics models, we additionally present all existing learning based models for rumor propagation in rumor source detection problem. It has been shown that data representation learning based approaches have shown promising results by modeling the information diffusion process in graphs that leverage both graph structure and node feature information. Although this approach results in the best precision performance among all baselines, time of training and tuning parameters still requires more investigations to be reduced. We notice that this research direction is growing rapidly now for rumor or misinformation source detection problems. Meanwhile, we also notice that there is little work for investigating more efficient and effective schemes for estimating rumor sources in the rumor source detection problem.    

Overall, we expect that our summarized and presented schemes in this paper will provide an overview of related works that can be applied for source detection problem in other domain specific networks such as wireless sensor network, network of virus spread, epidemic network, financial network and so forth.

\bibliography{ref} 

\end{document}